\documentstyle[preprint,aps]{revtex}

\draft

\begin{document}

\title{
Number of branches in diffusion-limited aggregates: The skeleton
}

\author{Stefan Schwarzer,$^{1,2}$  Shlomo Havlin,$^{1,3}$ Peter
Ossadnik,$^{1,4}$ and H. Eugene Stanley$^1$}

\address{$^1$ Center for Polymer Studies and Department of Physics,\\
Boston University, Boston, MA 02215}
\smallskip
\address{$^2$ Laboratoire de Physique M\'ecanique des Milieux
  H\'et\'erog\`enes, Ecole Sup\'erieure de Physique et Chimie Industrielles,\\
  75231 Paris, Cedex 05, France}
\smallskip
\address{$^3$ Department of Physics, Bar-Ilan University, Ramat-Gan, Israel}
\smallskip
\address{$^4$ Thinking Machines Corporation, c/o GMD, Schloss Birlinghofen,\\
  Postfach 1316, 53757 St. Augustin, Germany}

\date{\today}

\maketitle

\begin{abstract}

  We develop the skeleton algorithm to define the number of main
  branches $N_b$ of diffusion-limited aggregation (DLA) clusters.  The
  skeleton algorithm provides a systematic way to
  remove dangling side branches of the DLA cluster and has
  successfully been applied to study the ramification properties of
  percolation. We study the skeleton of comparatively large ($\approx
  10^6$ sites) off-lattice DLA clusters in two, three and four spatial
  dimensions. We find that initially with increasing distance from the
  cluster seed the number of branches increases in all dimensions.
  In two dimensions, the increase in the number of
  branches levels off at larger distances,
  indicating a fixed number of $N_b = 7.5\pm 1.5$
  main branches of DLA. In contrast, in three and four dimensions, the
  skeleton continues to ramify strongly as one proceeds from the
  cluster center outward, and we find no indication of a constant
  number of main branches. Likewise, we find no indication for a fixed
  $N_b$ in a study of DLA on the Cayley tree.  In two dimensions, we find
  strong corrections to scaling of logarithmic character,  which can
  help to explain recently reported deviations from self-similar
  behavior.

\end{abstract}


\narrowtext

\section{Introduction}
\label{s:skel-introduction}

The fractal nature of diffusion-limited aggregation has been
established by a number of studies, and a theoretical framework is being
developed~\cite{Witten81,Pietronero95}. In the initial paper written by
Witten and Sander \cite{Witten81}, power-law behavior of the two-point
correlation functions was found. Later, Meakin \cite{Meakin83a}
observed power-law scaling of the radius of gyration of DLA in two
(2D), three (3D), and four dimensions (4D).  Enhanced on- and
off-lattice algorithms facilitated studies of larger aggregates (Meakin
\cite{Meakin85}, Ball and Brady \cite{Ball85a}) and a striking
influence of the lattice structure on the structure of the aggregates
was found \cite{Ball85}.  Advanced off-lattice simulations by Ossadnik
\cite{Ossadnik93} grew clusters up to $M=5\times 10^7$. All these
studies indicate or confirm a fractal dimension of $d_f = 1.715 \pm
0.004 $ as reported by Tolman and Meakin\cite{Tolman89a}.  Apart from
the scaling of the radius of gyration with cluster mass, fractality is
revealed in studies of the branching structure of the aggregates by,
e.g., Alstr\o m {\it et al.} \cite{Alstrom88a}, Hinrichsen {\it et
  al.} \cite{Hinrichsen89}, Ossadnik \cite{Ossadnik92}, and more
recently Yekutieli {\it et al.}~\cite{Yekutieli94}. Self-similarity
has also been found in a study
by Argoul {\it et al.} calculating the moments of the mass
distribution of on-lattice clusters \cite{Argoul88a}, by Barab{\'a}si
and Vicsek's mapping of the cluster perimeter to a self-affine surface
\cite{Barabasi90a}, in investigations of the fjord geometry of DLA
\cite{Barabasi90,Schwarzer93}, and many more (for reviews, see,
e.g,
\cite{r:Meakin88,b:Bunde92,b:Feder88,b:Pietronero90,b:Stanley86,b:Vicsek92}).

However, almost as old as the model itself are reports about
deviations from a simple statistically self-similar structure. In
Refs.~\cite{Ball85a,Meakin87b} lattice anisotropy and deposition
habit effects on the structure of lattice clusters were studied,
Meakin and Vicsek \cite{Meakin85c} found that the tangential and
radial two-point-correlation functions in DLA display different
exponents. In recent works however,
Hegger and Grassberger~\cite{Hegger94} point out that off-lattice DLA
in strip geometry displays a slow approach to local isotropy.
Differences arise in the fractal dimensions measured in
circular or strip geometry \cite{Evertsz89}. Simple self-similar
scaling is not sufficient to fully characterize the mass distribution
within a DLA cluster, as recognized by Meakin and Havlin
\cite{Meakin87a}, Vicsek, Family and Meakin \cite{Vicsek90}, Amitrano,
Coniglio, Meakin, and Zanetti \cite{Amitrano91}, and Mandelbrot
\cite{Mandelbrot92}. Mandelbrot examined the
lacunarity~\cite{b:Mandelbrot82,Allain91} of off-lattice DLA and found
a trend towards increasing compactness for increasing systems size
which is not reflected in the measurements of the radius of gyration
of the clusters.

In our study we will apply the ``skeleton'' algorithm
\cite{Havlin84,Havlin85} to calculate the number of main branches
$N_b$ in off-lattice DLA (Sec.~\ref{s:skel-algorithm}). We obtain
the skeleton of a DLA cluster by removal of all the small dangling
side branches of the cluster (we define the skeleton in the following
section). Only the main branches that reach out to a certain fraction
of the cluster radius remain. For a self-similar, scale-invariant
structure we expect that the large-scale properties, like the number
of main branches of the cluster, are not altered under rescaling.
Conversely, we take an increase or decrease of the number of main
branches {\it with increasing cluster size} to suggest that the
structure is not self-similar.

In 2D, we find that significant finite size effects of logarithmic
nature are present which may be related to some of the observations
reporting deviations from simple self-similar behavior. Asymptotically
however, the number of main branches of 2D DLA approaches a constant
number (Sec.~\ref{s:skel-d2}). In contrast to the behavior in 2D, the
number of branches of 3D and 4D DLA increases with cluster mass,
indicating that these clusters may not be self-similar
(Sec.~\ref{s:skel-d-34}).

\section{The skeleton}
\label{s:skel-algorithm}

DLA has a loopless tree structure \cite{Meakin84}. This observation is
apparent in simulations of off-lattice aggregates
\cite{Tolman89,Ossadnik91} where an incoming particle is added to the
cluster when its distance to the cluster is below a specific sticking
distance. Its ``parent'' particle is the one that it was closest to at
the moment of incorporation into the cluster.  The child-parent
relationship allows us to uniquely assign a generation number or
``chemical distance from the seed'' to every cluster particle.  To
this end, we assign to the seed particle of the DLA cluster the number
$\ell=0$.  The children of the seed are assigned the chemical distance
$\ell=1$; children of $\ell=1$ particles have chemical distance
$\ell=2$. In general, each particle inherits the chemical distance of
its parent and increases this number by one to find its own.  The
chemical distance turns out to be a very useful quantity in studies of
the branching properties of DLA
\cite{Alstrom88a,Hinrichsen89,Ossadnik92} and we will use it here to
define the skeleton of DLA --- closely following Ref.~\cite{Havlin84}.

Roughly speaking, we obtain the skeleton by removal of all the small
dangling side branches of the cluster. To this end we first identify
all the sites that are tips of branches --- those sites to which no
other particle has been attached during the growth process. The $\ell$
value of each tip $\ell_{\rm tip}$ is then passed back to its parent and
grandparent and so forth until we hit an ancestor site which has more than one
child. In this case, only the largest of the $\ell_{\rm tip}$ values
of the children is retained. We iterate this process until all cluster
sites ``know'' their $\ell_{\rm tip}$ value.

Now let us choose an arbitrary value
$\ell_c$ less than the largest chemical distance $\Lambda$ of any site
from the seed. Given $\ell_c$, we define as the skeleton those sites
for which both $\ell_{\rm tip} \ge \ell_c$ and $\ell \le \ell_c$
holds. Thus, side branches which have not grown out to at least
$\ell_c$ do not contribute to the skeleton.

In Fig.~\ref{f:skel-skeleton}, we display the skeletons of two
randomly selected growing DLA clusters at mass
$M=5\,000,~50\,000,~500\,000$. In this figure, $\ell_c$ is chosen to
be half the ``chemical radius'' $\Lambda/2$ of the cluster.  Note that
the termination points of the skeleton are located almost on a circle,
although we use the chemical and not the Euclidean distance from the
seed to define the skeleton. This indicates that DLA grows radially
outward without forming loops. Indeed, it has been found that the
fractal dimension $d_{\rm min}$ of the path connecting the
cluster seed to a given site is equal to $1$ \cite{Meakin84}. Here,
$d_{\rm min}$ is defined by $\ell \sim r^{d_{\rm min}}$.
In other words, for
cluster sites at a given distance $r$ from the cluster seed, the
average value $\langle \ell \rangle$ is proportional to $r$, $\langle
\ell \rangle \sim r$. For the remainder of this study, we will
consider $\ell$ and $r$ as equivalent.

One physical interpretation of the skeleton in DLA can be obtained if we
consider the aggregate as a conductor situated between the grounded seed
and a circular electrode (a sphere or hypersphere in 3D or 4D,
respectively) of ``radius'' $\ell_c$.
Then the skeleton is the collection of paths that contribute to the
current through the aggregate.

The skeleton as defined above has some desirable and some less
desirable features. First, its definition is certainly simple, which
is desirable. Second, we see in Fig.  \ref{f:skel-skeleton} that in an
intermediate range of $\ell$ the number of branches in the skeleton is
small, which is another desirable feature since we claim that the
skeleton identifies the main branches of the DLA cluster. However, we
see that close to $\ell_c$ a lot of branches appear. Their presence
ultimately reflects our ignorance as to whether a specific side branch
will keep growing or die out. Thus, these branches are a very
``physical'' property, but, nevertheless, an undesirable feature. We
will address this problem by keeping $\ell_c$ a variable and
systematically test the dependence of the skeleton calculations on this
parameter.

We define $\lambda \equiv \ell_c /\Lambda$ and consider ensembles of
DLA clusters for which the the value $\Lambda$ is fixed. For such an
ensemble, we define $B(\lambda, \Lambda; \ell)$ as the average number
of sites in the skeleton at chemical distance $\ell$ from the cluster
seed. Note that $B(\lambda, \Lambda; \ell)$ is defined only for $\ell \le
\ell_c$.

Some general conclusions about properties of the
DLA skeleton can be drawn from the definition and are visible in Fig.
\ref{f:skel-skeleton}.
\begin{itemize}
\item[(i)] For $\ell =0$, only the cluster seed contributes, thus
  $B(\lambda,\Lambda;\ell=0) = 1$ independent of spatial dimension $d$
  and cluster size $\Lambda$.
\item[(ii)] Since $\ell\sim r$, and since DLA is essentially loopless so
  that overhangs can be neglected, we can determine the value
  $B(\lambda,\Lambda;\ell=\ell_c)$ from the number of intersection
  points of a ``cookie cutter'' of radius corresponding to $\ell_c$
  with the DLA cluster. As the analog of a cookie cutter in 3D we
  consider a sphere, and in 4D a hypersphere. The result is
  $B(\lambda,\Lambda;\ell_c) \sim \ell_c^{\rm d_f -1}$ where $d_f$
  denotes the fractal dimension of DLA, which depends on the embedding
  dimension $d$. The exponent $d_f-1$ follows from the fact that the
  codimension of the intersection of two sets is obtained as the sum
  of the codimensions of the two intersecting sets.
\item[(iii)] The number of sites in the skeleton increases
  monotonically with $\ell$, i.e., $B(\lambda,\Lambda;\ell') \ge
  B(\lambda,\Lambda;\ell)$ if $\ell' \ge \ell$. Qualitatively we
  observe from Fig.  \ref{f:skel-skeleton} that the increase is slow
  in the vicinity of the cluster seed, but is followed by a
  precipitous increase close to $\ell_c$.
\item[(iv)] For fixed $\ell$ and cluster ``radius'' $\Lambda,$ the
  number of sites in the skeleton decreases monotonically when
  $\lambda$ is increased, because some branches that contribute to
  the skeleton at some value of $\ell_c$ fail to reach larger
  values.  In the extreme case, when $\lambda=1$, we obtain
  $B(1,\Lambda;\ell) = 1$, independent of $\Lambda$ and $\ell$.
\end{itemize}

In the study of the skeleton of percolation clusters \cite{Havlin84},
a slow increase of the number of arms for small $\ell$ was found to be
due to finite size corrections. Taking this into account, one obtains
for small $\ell$ a {\it constant\/} number of branches. In analogy to
the percolation example, we use finite-size scaling techniques to
extract information about DLA from our 2D, 3D and 4D skeleton calculations,
which we present in the following sections.  The DLA results in finite
dimensions will be compared to the skeleton of a random Cayley tree
model, which can be considered to be the limit of infinite
dimensionality of DLA \cite{Vannimenus84}.

\section{Skeleton of 2D DLA}
\label{s:skel-d2}

In Fig. \ref{f:skel-2d} we display $B(\lambda=0.5, \Lambda;\ell)$ as a
function of $\ell/\ell_c$ for $\Lambda = 100,$ $316,$ $1\,000,$
$3\,162$, and $5\,000$. For $\Lambda \le 1\,000$, we obtain our data
from $500$ off-lattice clusters, for $\Lambda \ge 3\,162$, $114$
clusters are averaged. The clusters are grown stepwise and growth is
interrupted for the analysis when $\Lambda$ equals one of the above
listed values. For $\Lambda = 5\,000$, the cluster mass $M$ is
typically $M= 1- 1.5\times10^6$ sites.

As a function of $\ell/\ell_c$
we find that after an initial short transient the central part of the
data displays approximately linear behavior. The linear region widens
when the cluster size $\Lambda$ increases. The approximate linear
behavior terminates when $\ell/\ell_c$ approaches $\approx 0.3$.
Beyond $\ell/\ell_c \approx 0.3$, $B(0.5,\Lambda;\ell)$ increases
sharply towards the final value $\sim \ell_c^{d_f-1}$. However, the
linear region of the data tends to become flatter as we proceed to
larger $\Lambda$.  Since $B(\lambda,\Lambda)$ is monotonically {\it
  increasing,} the slope $\alpha(\lambda,\Lambda)$ in the linear
region cannot be negative. Thus, asymptotically
$\alpha(\lambda,\Lambda)$ can either approach $0$ or assume a finite
value $>0$. We will refer to the restricted range of approximately
linear behavior as the ``scaling regime'' of the skeleton.

In order to distinguish between these two possibilities, we examine the
dependence of $\alpha(\lambda,\Lambda)$ on system size, here
characterized by $\Lambda$. To this end, we determine the slope
$\alpha(\lambda,\Lambda)$ of a linear least squares fit,
\begin{equation} \label{e:skel-fit}
        \ln B(\lambda,\Lambda;\ell)\approx \alpha(\lambda,\Lambda)
            \ln (\frac{\ell}{\ell_c}) + \beta (\lambda,\Lambda),
\end{equation}
to $\ln B(\lambda,\Lambda;\ell)$ vs $\ln (\ell/\ell_c)$ for
data points satisfying $\ell \ge 10$ and $\ell/\ell_c \le 0.3$. We use
the assumption, familiar from the estimate of critical exponents from
finite system size calculations (see, e.g. \cite{b:Binder86}), that
$\alpha(\lambda,\Lambda)$ may be written as
\begin{equation}
\label{e:skel-alpha-finite-general}
        \alpha(\lambda,\Lambda) =
             \alpha_\infty (\lambda) + f(\lambda,\Lambda) + \dots.
\end{equation}
Here, the leading correction term $f(\lambda,\Lambda)$ to the asymptotic value
$\alpha_\infty (\lambda)$ is typically either (i) a power law $\sim
\Lambda^{-\gamma}$ or (ii) a logarithmic term, e.g., $\sim 1/\ln
\Lambda$. Both vanish for $\Lambda\to\infty$. The dots indicate
corrections that decrease faster than $f(\lambda,\Lambda)$ as $\Lambda
\to\infty$.

In Fig. \ref{f:skel-2d-finite}a we plot
$\alpha(\lambda;\Lambda)$ vs $1/\ln \Lambda$ for different values of
$\lambda$. From the almost linear behavior of $\alpha(\lambda;\Lambda)$
for large $\Lambda$ we conclude that the leading order correction term
can be approximated by the reciprocal of $\ln \Lambda$ (ii), i.e.,
\begin{equation} \label{e:skel-2d-correction-term}
        f(\lambda,\Lambda) = A(\lambda) / \ln \Lambda.
\end{equation}
Here, $A(\lambda)$ denotes an amplitude which depends on $\lambda$ as
evident from Fig.~\ref{f:skel-2d-finite}a. However, our calculations do not
allow us to {\it exclude\/} the possibility that $f(\lambda,\Lambda)$
vanishes in the power-law fashion (i), although, in this case
the characteristic power $\gamma$ must be small, $\gamma < 0.2$.
Larger values of $\gamma$ would lead to significant curvature in the
large $\Lambda$ behavior of $\alpha(\lambda,\Lambda)$ [cf. Fig.
\ref{f:skel-2d-finite}b].

{}From figure \ref{f:skel-2d-finite}a we conclude that $\alpha_\infty
(\lambda)= 0$ for $0.3 \le \lambda \le 0.8$ by extrapolating the trend
displayed by $\alpha(\lambda,\Lambda)$ for the $\Lambda$ range
accessed by our simulation. Since for $\lambda\to 1$ ,
$B(\lambda,\Lambda;\ell)= 1$ for all $\ell$, we expect this result to
also hold if $\lambda > 0.8$.  For small $\lambda$, the skeleton
probes the frozen region of the DLA cluster which effectively stopped
growing \cite{Amitrano91}, so that our results should be less effected
by possible slow structural changes in the growth zone of the cluster
\cite{Schwarzer93,Plischke84,Ossadnik92a}. Therefore, although the
fit region of $B(\lambda,\Lambda;\ell)$ becomes narrower, we assume
that $\alpha_\infty (\lambda)= 0$ extrapolates into the small
$\lambda$ regime.

The implications of $\alpha_\infty(\lambda) = 0$ are intriguing.
Asymptotically, we can assign a {\it fixed\/} number of main branches
$N_b$ to 2D DLA growth. We deduce the actual number of these branches
from the behavior of the intercepts $\beta(\lambda;\Lambda)$ of the
fit (\ref{e:skel-fit}) in the ``flat'' region of the skeleton. In Fig.
\ref{f:skel-2d-finite}c, $\beta(\lambda;\Lambda)$ is plotted as a
function of $1/\ln \Lambda$ for several $\lambda$.  The value of
$\beta(\lambda;\Lambda)$ slightly increases as the cluster size grows
--- corresponding to more main branches in larger systems. However,
the increase levels off at about $\beta_\infty (\lambda) = 2\pm 0.2$,
so that on average $N_b = \exp[\beta_\infty (\lambda)] = 7.5 \pm 1.5$
main branches result.

\section{DLA in higher dimensions: 3 and 4}
\label{s:skel-d-34}

We also perform simulations of off-lattice DLA in
3D and 4D. Our findings are quite different from the 2D case, a result
not surprising in the light of earlier work suggesting that 2D is a
special case~\cite{Schwarzer92a}. In
particular, our calculations suggest that in contrast to the 2D case, where
$\alpha_\infty(\lambda) = 0$, in the 3D and 4D case
$\alpha_\infty(\lambda) > 0.$ We discuss in the following the
numerical data supporting this result.

\subsubsection{Skeleton of 3D DLA}

In Fig.~\ref{f:skel-3d} we display $B(\lambda=0.5,\Lambda;\ell)$ for
3D DLA. Different curves are characterized by different values of
$\Lambda.$ The largest value of $\Lambda$ is $560$, which corresponds
to DLA clusters with $M \approx 10^6$ particles. The skeletons of
$450$ $(\Lambda \le 200)$ and $75$ $(\Lambda > 200)$ clusters have
been averaged to obtain the displayed data.

The tendency for the central part of the plot to become flatter as
$\Lambda$ increases is much less pronounced than in 2D. This
observation is supported by Fig.~\ref{f:skel-3d}a, which shows the
$\Lambda$ dependence of $\alpha(\lambda;\Lambda)$. We obtain the
slope $\alpha(\lambda;\Lambda)$ from a linear least square fit of $\ln
B(\lambda,\Lambda;\ell)$ in the region characterized by $\ell\ge 10$
and $\ell/\ell_c < 0.3.$ Like in 2D, also in 3D the correction term is
$\sim 1/\ln\Lambda$
(\ref{e:skel-alpha-finite-general}), and likewise we cannot exclude
that the correction is of power-law type with a small exponent.
However, unlike in 2D, in the 3D case the slopes corresponding to the
small $\lambda$ values associated with the ``frozen'' region of the
cluster $\alpha(\lambda;\Lambda)$ extrapolate to a non-zero value,
$\alpha_\infty(\lambda)\approx 0.2$. This value indicates that in 3D
the number of branches of the skeleton increases monotonically in
power-law fashion as a function of $\ell/\ell_c$ even in the limit of
infinite cluster size.

\subsubsection{Skeleton of 4D DLA}

The skeletons of $50$ off-lattice DLA clusters were averaged to obtain
the data displayed in Fig.~\ref{f:skel-4d}. The clusters with $\Lambda
=190$ contain up to $\approx 500\,000$ particles.  Similar to the 3D
case, the slopes in the central part of the $B(\lambda,\Lambda;\ell)$
data do not tend to become flat with increasing mass and this behavior
is reflected in the $\Lambda$ dependence of $\alpha(\lambda;\Lambda)$,
which we display in Figs.~\ref{f:skel-4d-finite}a and b. The quality
of the 4D calculations is much poorer than that of the 3D and 2D
calculations. A finite value $\alpha_\infty (\lambda)$ is consistent
with the two finite size scaling plots Figs.~\ref{f:skel-4d-finite}a
and b.  As in 3D, such a finite value indicates a power-law branching
of the skeleton in the scaling regime as the cluster mass goes to
infinity.

Moreover, if we consider the intersections $\beta(\lambda,\Lambda)$ of
the straight line fits to the scaling region we observe that they may
diverge as $\Lambda \to\infty.$ Such behavior would imply that
$B(\lambda,\Lambda; \ell)$ may not converge to a limit as
$\Lambda\to\infty.$ Thus we expect rather strong deviations from
self-similar behavior in 4D and for $\Lambda\to\infty$ no self-similar
limit as in the 2D case may exist.

\section{Cayley tree model}

The realization of the DLA growth model on a Cayley tree is considered
to be the mean-field or infinite dimensionality limit of DLA. DLA on
the Cayley tree has been formulated as follows~\cite{Vannimenus84}.
To the cluster seed (shell $0$), we links $z$ surface sites (shell
$1$) --- $z \ge 2$ is the functionality of the model. In each growth
step, we select (occupy) one of the surface sites randomly. We then
create $z-1$ new (empty) surface sites and link them to the just
occupied site.  The shell number of the new empty sites is equal to
the shell number of the occupied parent incremented by $1$.  Note that all
occupied sites have $z$ links to neighbor sites and that all empty
sites have exactly $1$ neighbor site.

Since new cluster sites are randomly picked from the set of all
surface sites of the growing tree, we effectively disregard screening
of the interior parts of the tree, which in finite dimensions is the
essential mechanism to produce low dimensional fractal structures
\cite{Ball84}.  We have numerically grown clusters of this tree model
with functionality $z=3$, $4$ and $5$ as large as $32$ shells.  As a
function of the shell number, the number of occupied sites grows
exponentially \cite{Vannimenus84}. Similar to the case of 3D and 4D
DLA, the skeleton shows no sign of saturation as a function of
$\ell/\ell_c$.  The increase seems to have even exponential character, with
$\ln B(\lambda,\Lambda=32;\ell) \propto \ell/\ell_c$, and prefactors
that depend on $z$ and $\lambda$.

This behavior is consistent with the situation in 3D and 4D DLA, where
the skeleton as a function of $\ell/\ell_c$ also increases, albeit
in a power-law fashion.

\section{Discussion}
\label{s:skel-discussion}

\subsection{Previous results on the number of stable branches of DLA}
\label{s:skel-ball}

The number of branches $N_b$ of DLA has been used as an important
morphological characterization of the
cluster~\cite{Meakin85,Turkevich85,Meakin86,Ball86,Halsey86,Argoul90,Arneodo92}.
In 2D using conformal mapping arguments, Ball \cite{Ball86} has
proposed that the maximum number of stable branches $N_{\rm max}$ is
related to the mass scaling dimension $D$ of the cluster by
\begin{equation}
\label{e:n_of_d}
(N_{\rm max}/2 -1)(D-1) = 1.
\end{equation}
For $D \approx 1.7$ this relation predicts $N_{\rm max} \approx 4.9.$
Similarly, Turkevich and Scher~\cite{Turkevich85} have modeled the
the tips of the cluster as the corners of a convex polygonal cluster
envelope. The number $N$ of branches then determines the tip angles
and those in turn the mass scaling dimension, namely, $D-1 = N /
(N+2).$ If one replaces $D-1$ from Eq. (\ref{e:n_of_d}) and assumes
$N=N_{\rm max}$, we obtain $N_{\rm max} \approx 4.8$, in agreement
with the previous prediction.

Both numbers seem to be significantly smaller than our finding
$N_b 7.5 \pm 1.5$ in 2D.\footnote{
Concerning the Turkevich-Scher assumption of
a convex polygonal cluster envelope one may argue that it is possible
to conceive cluster envelopes with more acute angles and a larger
number of arms --- possibly around $7.$
}

It is not clear why $N_b$ should be different from $N_{\rm max}$
following from Eq.  (\ref{e:n_of_d}). On the one hand, if the actual
number of branches is larger than the maximum number of stable ones
then the competition among branches for the incoming flux of random
walkers will cause branches to die (cf. \cite{Halsey92}). If, on the
other hand, the number of branches is too small, then new branches
will be created because the fjords of the aggregate are not
sufficiently screened to suppress growth. If, in fact, $N_{\rm max}$
and $N_b$ do not coincide, the reason could be that large branches may
finally split, and one of the now too numerous arms will finally die
out, but slowly enough to increase the number of branches apparent in
the skeleton.

%
%

\subsection{Relation to ``lacunarity'' measurements on DLA}


Recently, Mandelbrot has investigated the lacunarity of off-lattice
DLA \cite{Mandelbrot92}. To calculate the lacunarity of a cluster of
span $\Lambda$, we (i) select the frozen interior region of
the cluster (of span $k\Lambda,$~$k<1$) and rescale all the particles
coordinates into the unit circle, then (ii) center solid disks of
radius $\epsilon \ll 1$ on the rescaled particle positions and (iii)
determine the area fraction $f$ of the unit circle covered by the
$\epsilon$ disks.  For a ``normal'' fractal, one expects $f$ to
saturate when $\Lambda$ increases and in this case to provide a measure for
the lacunarity of the object. For DLA however, $f$ increases
continuously with $\Lambda$. Mandelbrot argues accordingly that either DLA
displays a ``massive transient'' or a ``limitless drift'' towards
increasing compactness.

We argue that the logarithmic finite size corrections that we see
in 2D in the approach of $B(\lambda,\Lambda;\ell)$ to its asymptotics
can explain the ``massive transient'' picture. As we have seen above,
$B(\lambda,\Lambda;\ell)$ increases with system size corresponding
to structurally different clusters with more main branches
$N_b$. Consequently, the area fraction $f$ increases.

Thus, off-lattice DLA shows a very slow structural change towards its
asymptotic structure, similar to its on-lattice siblings.

\subsection{Why is 2D DLA different from higher dimensional DLA?}
\label{s:skel-2D}

The remaining open question is why on the one hand $\alpha_\infty
(\lambda) \to 0$ in 2D, but, on the other hand, $\alpha_\infty
(\lambda) > 0$ in 3D and 4D and in the latter case possibly even a
divergence of $\beta(\lambda,\Lambda)$ with $\Lambda$.  Let us first
note, that 2D is known to be a ``critical dimension'' for radial DLA
with respect to the scaling properties of the growth probabilities of
the surface sites~\cite{Schwarzer92a,Stanley93,Stanley94}.  We find
two more indications supporting our result, one from a simple
consideration of the cluster density in different spatial
dimensionalities, the other from a comparison of the correlated DLA
tree to a random tree model, for which the scaling behavior of the
skeleton is known \cite{Havlin85}.

\subsubsection{Cluster ``density'' and screening}
\label{s:skel-cluster-density}

We have already noted above that the number of branches in the
skeleton at its termination points at $\ell_c$ is $\sim \ell_c^{d_f
  -1}.$ Thus, for given $\ell=\ell_c$, there are many more branches
present in the 3D and 4D skeleton than in the 2D skeleton.  In order
that these branches contribute to the skeleton in the scaling region
they must originate deep in the cluster interior.

With increasing dimension, the screening of the cluster interior from
incoming particles decreases. This decrease manifests itself for
example in the strong semi-exponential screening of the cluster fjords
in 2D \cite{Lee88,Schwarzer90,Mandelbrot91,Wolf91,Wolf93} in contrast to the
power-law type \cite{Schwarzer92a} screening in 3D or in the screening
free Cayley-tree model --- where we also observe $\alpha_\infty
(\lambda) > 0$. Qualitatively, the reduction of screening
can also be seen in the mean-field
expression for the particle penetration depth $\sim \rho^{-1} \sim
\Lambda^{d-d_f}.$ The penetration depth increases with $d$, because $d_f$
approaches $d-1$ from above.\footnote{
It is however known from simulations that at least in 2D this relation
does not hold and the penetration depth is $\sim \Lambda$.
}

Therefore we conclude that the weaker screening in dimensions $d>2$
is responsible for the further increase in the number of branches of
the skeleton both as a function of $\ell/\ell_c$ and as a function of
cluster size $\Lambda$.

\subsubsection{Comparison to a random tree model}
\label{s:skel-random-tree}

DLA has the structure of a tree.
Thus, we find it instructive to compare our findings for the off-lattice
DLA skeleton to results for the skeleton of a {\it random} tree model
\cite{Havlin85}. This tree model is defined on a 2D square lattice and
characterized by a tunable ``intrinsic'' dimension $d_\ell$, which
determines how the mass $M(\ell)$ of the tree object depends on the
chemical distance $\ell$ from the cluster seed, i.e., $M(\ell) \propto
\ell^{d_\ell}.$ To this end, we occupy randomly exactly $N(\ell) =
\ell^{d_\ell - 1}$ sites in the $\ell$th chemical shell (or fewer if
necessary) --- all the other sites are blocked. Only those sites in
the $\ell$th shell shall be occupied that do not close loops, i.e., that
are not neighbors to sites in previous shells. Note that apart
from possible constraints due to the embedding lattice at small
$\ell$, the {\it topological\/} structure of the clusters does {\it not\/}
depend on the dimension of the embedding lattice.

Reference \cite{Havlin85} then considers the skeleton
$B(1,\Lambda;\ell)$ of a tree with intrinsic dimension $d_\ell$. In
contrast to DLA, here in general $B(1,\Lambda;\ell) \ge 1$, because
the last chemical shell contains $N(\Lambda) = \ell^{d_\ell -1} \gg 1$
sites. In these random tree models there exists a sharp transition
value $d_\ell^c = 1.65\pm 0.05$ which separates two regions with
different types of behavior.  For trees with $d_\ell < d_\ell^c$ the
skeleton displays a flat --- approximately constant --- region as a
function of $\ell$, like the one observed in the 2D DLA case ({\it
  non-branching\/} skeleton). However, for larger values of $d_\ell$
the skeleton displays a well defined power-law increase as function of
$\ell,$ i.e., $\alpha_\infty(1) > 0$ ({\it branching\/} skeleton).

Thus, if we consider DLA as such a random tree, we expect a branching
skeleton for spatial dimension $d>2$, since $d_\ell = d_f > d_\ell^c.$
In 2D, $d_f$ is only slightly larger than $d_\ell^c$, or even equal to
$d_\ell^c$ within the error bars, which is consistent with the
observed non-branching skeleton.

\section{Summary}

We have determined the skeleton of comparatively large ($\approx
10^6$ sites) off-lattice DLA clusters in 2D, 3D and 4D.  We find that,
asymptotically, in 2D the skeleton of DLA suggests a fixed number of
$N_b\approx 7.5\pm 1.5$ main branches and a self-similar structure.
In 3D and 4D, and possibly all spatial dimensions $d>4$, the DLA
skeleton is a ramified object, which displays branching over the whole
range of $\ell$-values for which it is defined.

For all dimensions we find strong finite-size effects corresponding to
a slow change in the structure of DLA as the cluster size increases.
The presence of strong corrections to scaling of logarithmic character
in 2D DLA is in agreement with findings of deviations from
self-similar behavior reported elsewhere.

Moreover, in 4D, it is possible that the structure of the aggregate
keeps changing even asymptotically, such that there is no self-similar
limit.

\section*{Acknowledgments}

We are grateful for enlightening discussions with Z.~Alexandrowicz,
A.-L.~Barab{\'a}si, S.V.~Buldyrev, A.~Coniglio, M.F.~Gyure,
U.~Essmann, H.J.~Herrmann, J.~Lee, M.~Ossadnik, G.~Peng, S.~Prakash
and S.~Sastry. We also kindly acknowledge financial support by the NSF
and benefits from computational resources at the HLRZ, KFA J\"ulich,
Germany.  S.S. is grateful to the scientific council of the NATO for
financial support (granted through the DAAD, Bonn).



\begin{figure}[hpbt]
%
%
\caption[Skeletons for 2 2D DLA clusters]{
\label{f:skel-skeleton}
Skeleton for 2 growing DLA clusters: (a--c) first cluster, (d--f)
second cluster. Growth of the cluster has been interrupted at cluster
masses $M=5\,000$ (a,d), $50\,000$ (b,e) or $500\,000$ (c,f). The
determination of the skeleton is based on a value of $\ell_c/\Lambda =
0.5$. The skeleton is then rescaled so that independent of cluster
mass the same size results.  }

\end{figure}


\begin{figure}[hpbt]
%
%
\caption[Skeleton $B(\lambda, \Lambda; \ell)$ for 2D DLA]{
\label{f:skel-2d}
Log-log plot of the 2D skeleton $B(\lambda=0.5,\Lambda; \ell)$
vs $\ell/\ell_c$ for
different $\Lambda$ values as indicated in the legend. For
$\Lambda \le 1\,300$ data are averaged over $850$ off-lattice DLA, for
$\Lambda > 1\,300$ averages over $114$ clusters are taken. For
comparison, we have also averaged data in the constant $M$ ensemble,
instead of the constant $\Lambda$ ensemble used here, and obtain
similar results (not shown). Note that in an intermediate region of
the plot ($\ell/\ell_c < 0.3$) a slope can be associated with
$B(\lambda=0.5,\Lambda; \ell)$ vs. $\ell/\ell_c$. As $\Lambda$ increases,
this slope becomes increasingly smaller, indicating that the asymptotic
curve may be flat in this ``scaling'' region.
Inset: The number of branches in the skeleton at $\ell = \ell_c$,
$B(\lambda, \Lambda; \ell_c)$ is equal to  the number of intersection
points of a cookie cutter of radius $\ell_c$ in $\ell$ space with the
cluster, i.e., $B(\lambda,\Lambda;\ell_c) \sim \Lambda^{d_f-1}.$ The
solid line is a guide to the eye and indicates a slope of
$d_f-1=0.70.$ Different symbols refer to different $\lambda = 0.8
(\bigcirc)$, $0.6 (\Box)$, $0.5 (\bigtriangleup)$, $0.4
(\bigtriangledown)$, $0.2 (\bullet).$
}
\end{figure}


\begin{figure}[hpbt]
%
%
\caption[Finite size scaling plots 2D DLA skeleton]{
\label{f:skel-2d-finite}
Finite size behavior of the slopes $\alpha(\lambda,\Lambda)$ of
$B(\lambda=0.5,\Lambda; \ell)$ vs $\ell/\ell_c$ calculated from data
in the range $\ell>10$ and $\ell/\ell_c < 0.3$.  The slopes
$\alpha(\lambda,\Lambda)$ for different $\lambda$ as indicated in the
legend of the figure are plotted vs (a) $1/\ln \Lambda$ and (b) $1/
\Lambda$. In (a) the slopes lie on asymptotically straight lines
indicating a finite size correction term $\sim 1/\ln \Lambda$. This
behavior should be contrasted to the calculations in (b) where the
correction term is assumed to be proportional to $1/L$ and pronounced
curvature is displayed as $\Lambda$ increases.  (c) Intercepts
$\beta(\lambda,\Lambda)$ of the straight line fit with the ordinate vs
$1/\ln \Lambda$. If $\alpha_\infty(\lambda) = 0$, then $\beta_\infty
(\lambda)$ denotes the logarithm of the number of arms $N_b$ in the
``flat'' region of the skeleton.  From $\beta_\infty\approx 2\pm 0.2$
we conclude that $N_b = 7.5 \pm 1.5$.}
\end{figure}


\begin{figure}[htbp]
%
%
\caption[Skeleton $B(\lambda, \Lambda; \ell)$ for 3D DLA]{
\label{f:skel-3d}
Log-log plot of the 3D skeleton $B(\lambda=0.5,\Lambda; \ell)$ vs
$\ell/\ell_c$ for different $\Lambda$ values as indicated in the
legend.  We take averages over $75$ ($\Lambda > 200$) and $450$
($\Lambda \le 200$) clusters. Note that the tendency of the power-law
region in the plot to become flatter with increasing $\Lambda$ is much
less pronounced than in the 2D case \protect\ref{f:skel-2d} and is in
fact rather indicating a {\it non-zero\/} $\alpha_\infty(\lambda)> 0$.
Inset: Number of branches in the skeleton at $\ell=\ell_c$. We expect
$B(\lambda,\Lambda;\ell_c) \sim \Lambda^{d_f-1}.$ The solid line is a
guide to the eye and indicates a slope of $1.5 \approx d_f -1$ in 3D.
Different symbols refer to different $\lambda$ and are the same as in
Fig.~\protect\ref{f:skel-2d}.}

\end{figure}


\begin{figure}[htbp]
%
\caption[Finite size scaling plots for 3D DLA skeleton]{
\label{f:skel-3d-finite}
Finite size behavior of the slopes $\alpha(\lambda,\Lambda)$ of
$B(\lambda=0.5,\Lambda; \ell)$ vs $\ell/\ell_c$
calculated from data in the range $\ell>10$ and $\ell/\ell_c < 0.3$ for
3D DLA.
The slopes $\alpha(\lambda,\Lambda)$ for different $\lambda$ with
symbols as in Fig. \ref{f:skel-2d-finite} are plotted vs (a) $1/\ln
\Lambda$ and (b) $1/\Lambda$.
(c) Intercepts $\beta(\lambda,\Lambda)$ of the straight line fit with
the ordinate vs $1/\ln \Lambda$.
}
\end{figure}


\begin{figure}[htbp]
%
\caption[Skeleton $B(\lambda, \Lambda; \ell)$ for 4D DLA]{
\label{f:skel-4d}
Log-log plot of the 4D skeleton $B(\lambda=0.5,\Lambda; \ell)$ vs
$\ell/\ell_c$ for different $\Lambda$ values as indicated in the legend.
As in the 3D case, the central region of the plot does not display a
tendency to become flat as $\Lambda$ increases. However, the system
sizes seem still too small to draw a final conclusion, in particular if
one considers the size dependence of the slopes in the power-law region
and the comparatively poorer quality of the calculations
(see Fig. \ref{f:skel-4d-finite}).
Inset: Number of branches in the
skeleton at $\ell=\ell_c$. We expect $B(\lambda,\Lambda;\ell_c)
\sim \Lambda^{d_f-1}.$ The solid line is a guide to the eye and
indicates a slope of $2.2 \approx d_f -1$ in 4D. Different symbols
refer to different $\lambda$ and are the same as in Fig.
\ref{f:skel-2d}.
}
\end{figure}


\begin{figure}[htbp]
%
%
\caption[Finite size scaling plots for 4D DLA skeleton]{
\label{f:skel-4d-finite}
Finite size behavior of the slopes $\alpha(\lambda;\Lambda)$ of
$B(\lambda=0.5,\Lambda; \ell)$ vs $\ell/\ell_c$
calculated from data in the range $\ell>10$ and $\ell/\ell_c < 0.3$ for
4D DLA.
The slopes $\alpha(\lambda,\Lambda)$ for different $\lambda$ with
symbols as in Fig. \ref{f:skel-2d-finite} are plotted vs (a) $1/\ln
\Lambda$ and (b) $1/\Lambda$.
(c) Intercepts $\beta(\lambda,\Lambda)$ of the straight line fit with
the ordinate vs $1/\ln \Lambda$.
}
\end{figure}

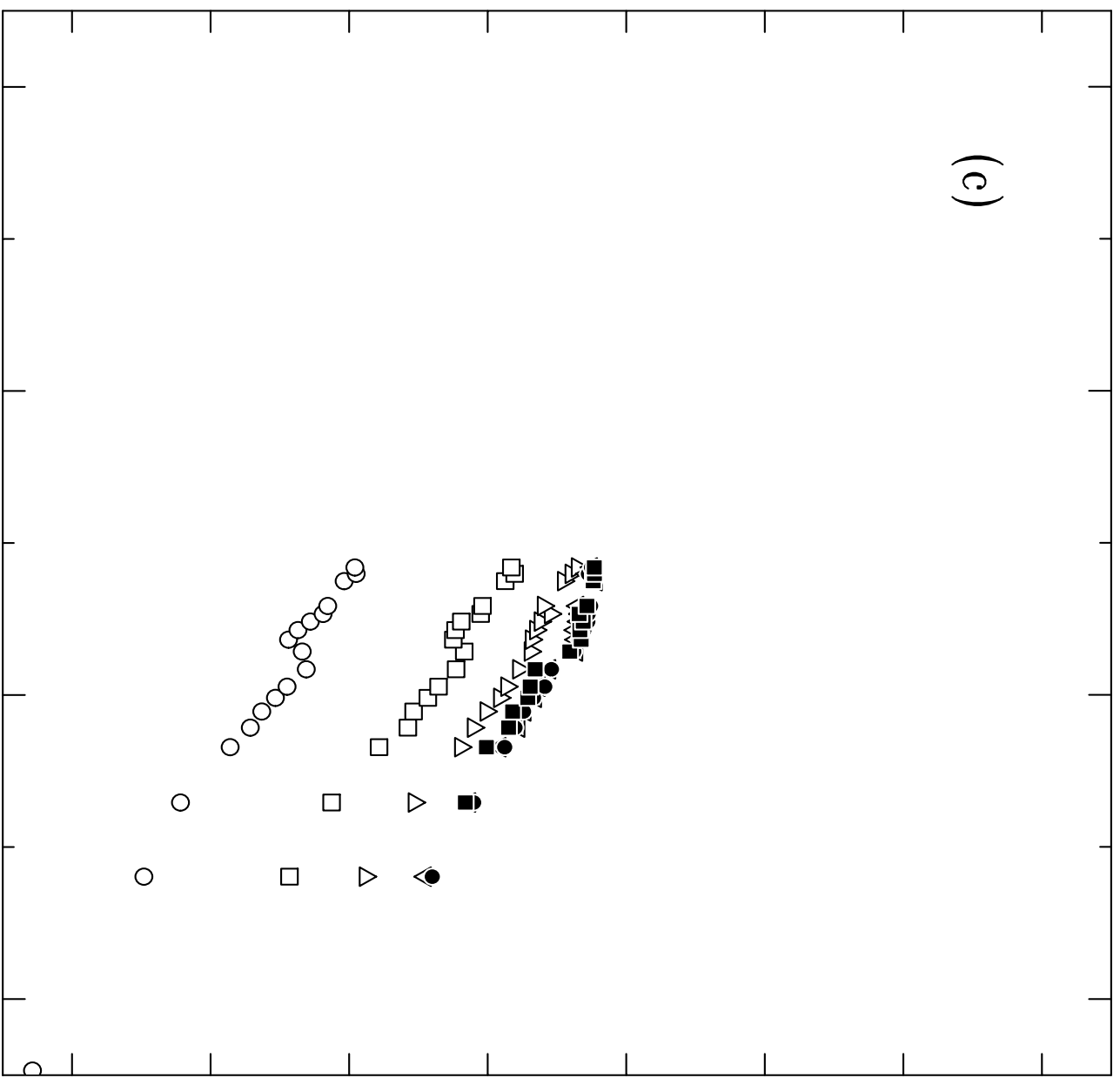
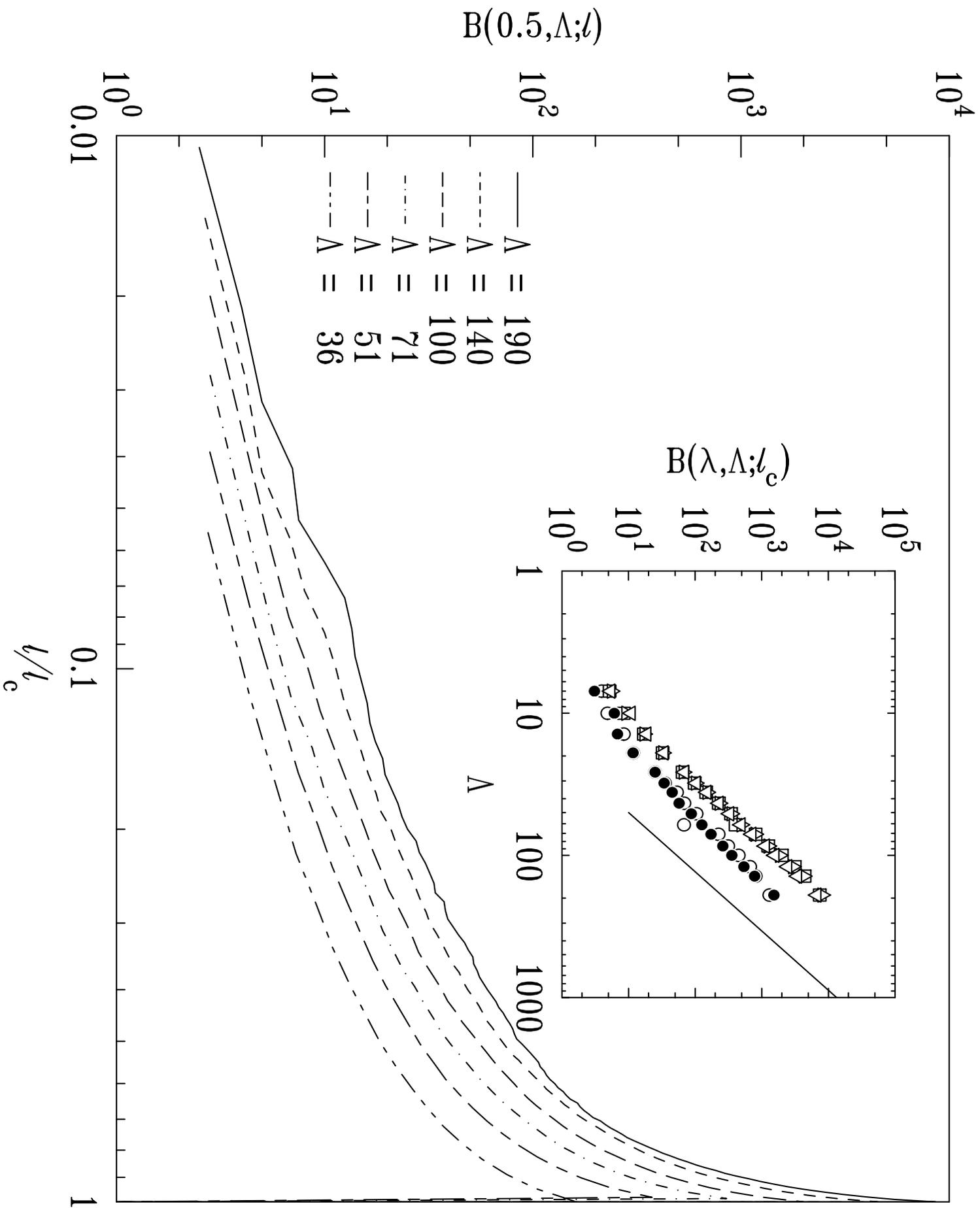
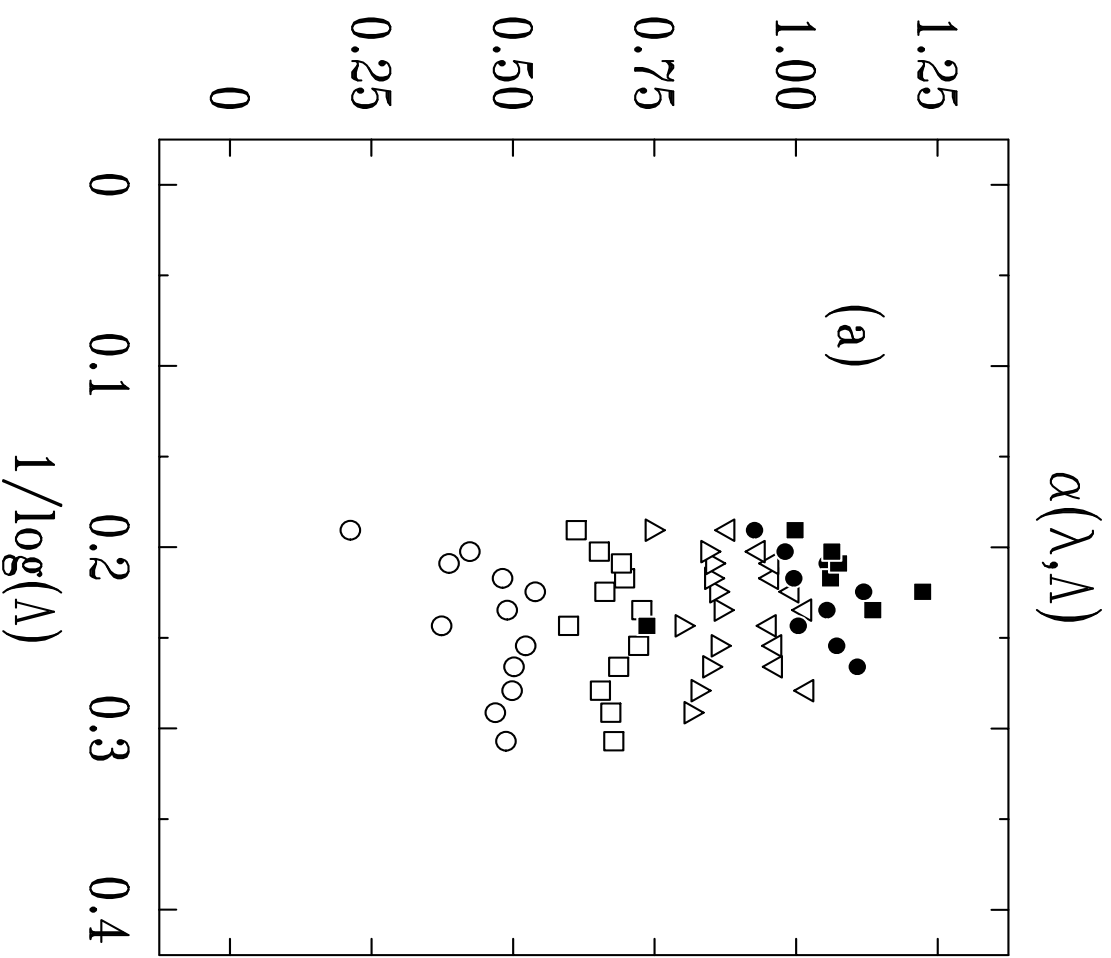
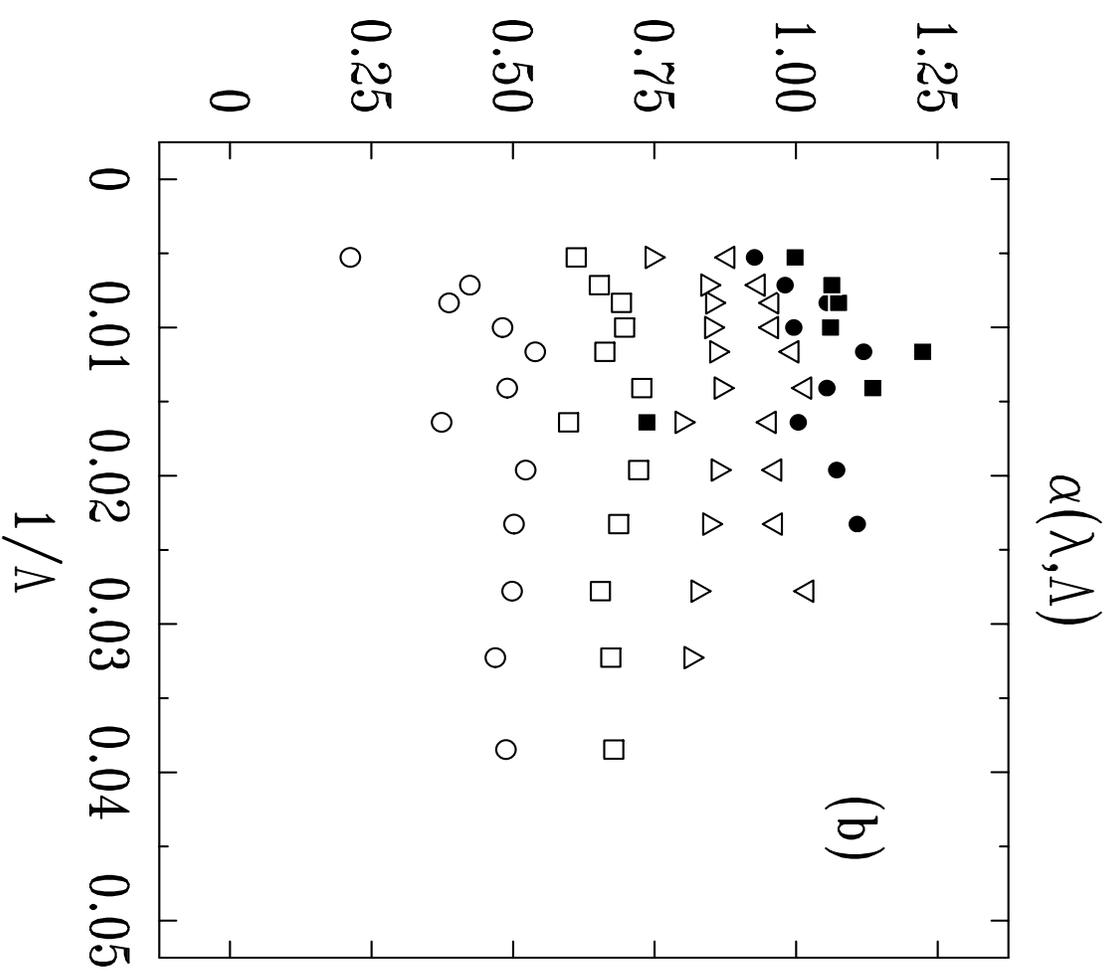
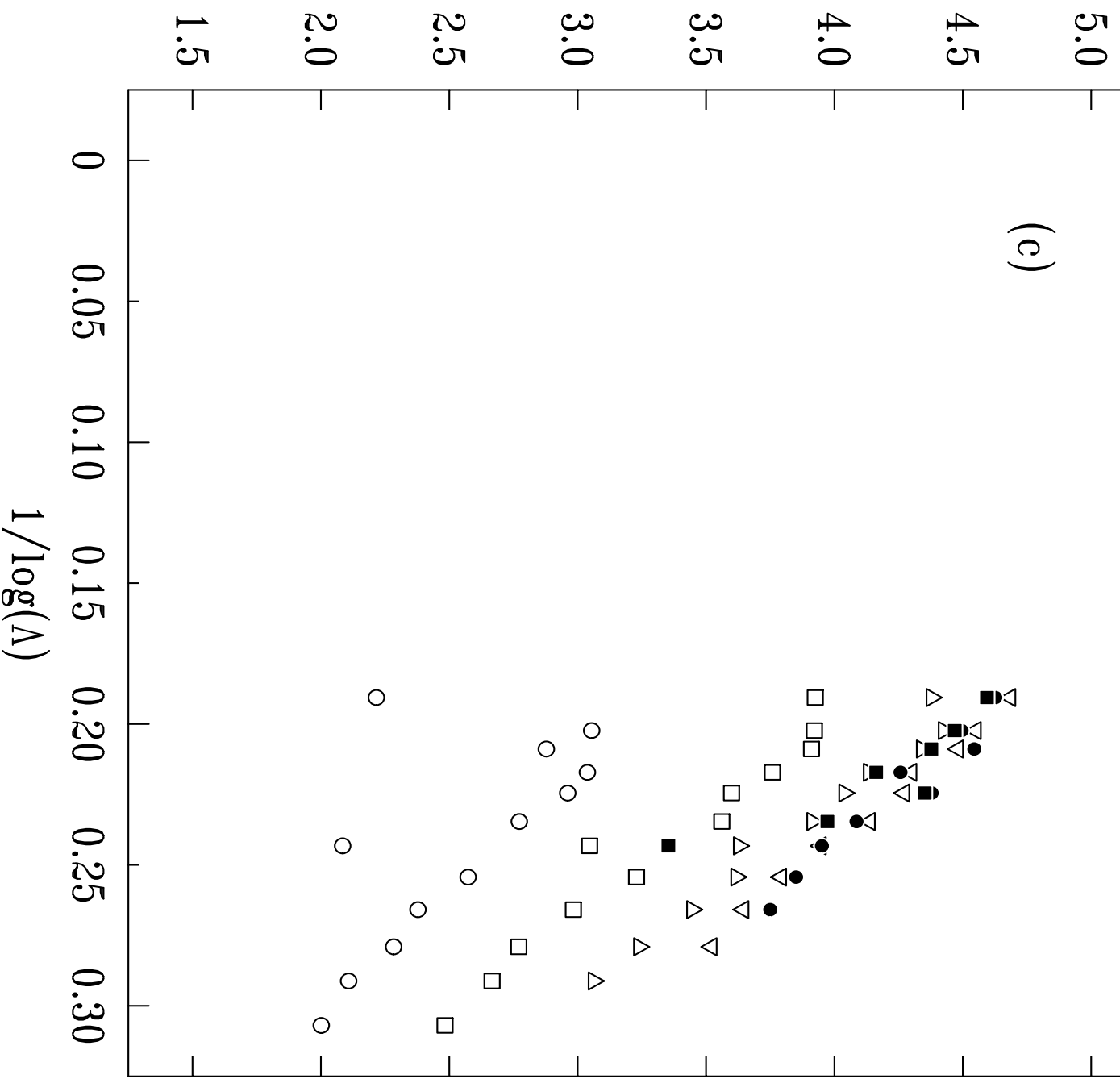
\end{document}